# vol2Brain: A new online Pipeline for whole Brain MRI analysis

José V. Manjón[1], José E. Romero[1], Roberto Vivo-Hernando[2], Gregorio Rubio[3], Fernando Aparici[4], Mariam de la Iglesia-Vaya[5,6] and Pierrick Coupé[7]

[1] Instituto de Aplicaciones de las Tecnologías de la Información y de las Comunicaciones Avanzadas (ITACA), Universitat Politècnica de València, Camino de Vera s/n, 46022, Valencia, Spain

[2] Instituto de Automática e Informática Industrial, Universitat Politècnica de València, Camino de Vera s/n, 46022, Valencia, Spain.

[3] Departamento de matemática aplicada, Universitat Politècnica de València, Camino de Vera s/n, 46022 Valencia, Spain.

[4] Área de Imagen Medica. Hospital Universitario y Politécnico La Fe. Valencia, Spain

[5] Unidad Mixta de Imagen Biomédica FISABIO-CIPF. Fundación para el Fomento de la Investigación Sanitario y Biomédica de la Comunidad Valenciana - Valencia, Spain.

[6] CIBERSAM, ISC III. Av. Blasco Ibáñez 15, 46010 - València, Spain

[7] CNRS, Univ. Bordeaux, Bordeaux INP, LABRI, UMR5800, PICTURA, F-33400 Talence, France

**\*Corresponding author:** José V. Manjón. Instituto de Aplicaciones de las Tecnologías de la Información y de las Comunicaciones Avanzadas (ITACA), Universidad Politécnica de Valencia, Camino de Vera s/n, 46022 Valencia, Spain.

Tel.: (+34) 96 387 70 00 Ext. 75275 Fax: (+34) 96 387 90 09.

E-mail address: jmanjon@fis.upv.es (José V. Manjón)

**Abstract**

Automatic and reliable quantitative tools for MR brain image analysis are a very valuable resources for both clinical and research environments. In the last years, this field has experienced many advances with successful techniques based on label fusion and more recently deep learning. However, few of them have been specifically designed to provide a dense anatomical labelling at multiscale level and to deal with brain anatomical alterations such as white matter lesions. In this work, we present a fully automatic pipeline (vol2Brain) for whole brain segmentation and analysis which densely labels (N>100) the brain while being robust to the presence of white matter lesions. This new pipeline is an evolution of our previous volBrain pipeline that extends significantly the number of regions that can be analyzed. Our proposed method is based on a fast multiscale multi-atlas label fusion technology with systematic error correction able to provide accurate volumetric information in few minutes. We have deployed our new pipeline within our platform volBrain (www.volbrain.upv.es) which has been already demonstrated to be an efficient and effective manner to share our technology with users worldwide.

# 1. Introduction

Quantitative brain image analysis based on MRI has become more and more popular over the last decade due to its high potential to better understand subtle changes in the normal and pathological human brain. The exponential increase of the current neuroimaging data availability and the complexity of the methods to analyze them make necessary the development of novel approaches able to address challenges related to the new "Big Data" paradigm (Van Horn and Toga, 2014). Thus, automatic, robust and reliable methods for automatic brain analysis will have a major role in near future, most of them powered by cost-effective cloud-based platforms.

Specifically, MRI brain structure volume estimation is being increasingly used to better understand the normal brain evolution (Coupe el al., 2017) or the progression of many neurological pathologies such as multiple sclerosis (Commowick et al., 2018) or Alzheimer's disease (Coupe et al., 2019) for example.

The quantitative estimation of the different brain structure volumes requires automatic, robust and reliable segmentation of such structures. Since manual delineation of the full brain is unfeasible for routine brain analysis (this task is too tedious, time consuming and prone to reproducibility errors) many segmentation methods have been proposed over the years. Some of them were initially focused at tissue level such as the famous SPM (Asburner and Friston, 2005). However, this level of detail may be insufficient to detect subtle changes in specific brain structures at early stages of the disease.

For example, hippocampus and lateral ventricle volumes can be used as early biomarkers of Alzheimer disease. At this scale, also cortical and subcortical grey matter structures are of special interest for the neuroimaging community. Classic neuroimaging tools such as the well-known FSL package (Jenkinson et al, 2012) or Freesurfer (Fischl et al., 2002) have been widely used over the last 2 decades. More recently, multi-atlas label fusion segmentation techniques have been extensively applied thanks to their ability to combine multiple atlas information minimizing mislabeling due to inaccurate registrations (Coupe et al., 2011, Wang and Yushkevich, 2013, Manjón et al., 2014, Romero et al., 2015).

However, to segment the whole brain in a large number of structures is still a very challenging problem even for modern multi-atlas based methods (Wang and Yushkevich, 2013; Cardoso et al, 2015; Ledig et al., 2016). First, the need of a large set of densely manually labeled brain scans and second the large amount of computational time needed to combine all those labeled scans to produce the final segmentation.

Fortunately, a fast framework based on collaborative patch-matching was recently proposed (Guiraud et al. 2016) to reduce the computational time required by multi-atlas patch-based methods.

More recently, deep leaning methods have been also proposed for brain structure segmentation. Those methods are mainly patch-based (Wachinger et al., 2018) or 2D (slice-based) (Roy et al., 2019) due to current GPU memory limitations. Current state-of-the-art whole brain deep learning based methods are based on ensembles of local neural networks such as the SLANT method (Huo et al., 2019) or more recently the Assemblynet method (Coupe et al., 2020).

The aim of this paper is to present a new software pipeline for whole brain analysis that we have called vol2Brain. It is based on an optimized multi-atlas label fusion scheme that has a reduced execution time thanks to the use our fast collaborative patch-matching approach and has been specifically designed to deal with both normal appearing and lesioned brains (a feature that most of preceding methods ignored). This pipeline automatically provides volumetric brain information at different scales in a very simple manner through a web-based service not requiring any installation or technical requirements in a similar manner as previously done by our volBrain platform that since 2015 has processed online more than 360,000 brains worldwide. In the following sections, the new pipeline will be described and some evidences of its quality will be presented.

## 2. Material and Methods

### 2.1. Dataset description

In our proposed method, we used an improved version of the full Neuromorphometrics dataset (http://www.neuromorphometrics.com) which consists of 114 manually segmented brain MR volumes corresponding to subjects with ages covering almost the full lifespan (from 5 to 96 years old). Dense neuroanatomical manual labeling of MRI brain scans was performed at Neuromorphometrics, Inc., following the methods described in (Caviness et al., 1999).

The original MRI scans were obtained from the following sources: 1) the Open Access Series of Imaging Studies (OASIS) project web site (http://www.oasis-brains.org/) (N=30), 2) the Child and Adolescent NeuroDevelopment Initiative (CANDI)

Neuroimaging Access Point (http://www.nitrc.org/projects/candi_share) (N=13), 3) the Alzheimer's Disease Neuroimaging Initiative (ADNI) project web site (http://adni.loni.usc.edu/data-samples/access-data/) (N=30), 4) the McConnell Brain Imaging Centre (http://www.bic.mni.mcgill.ca/ServicesAtlases/Colin27Highres/) (N=1) and 5) the 20Repeats dataset (http://www.oasis-brains.org/) (N=40).

Before manual labeling all the images were preprocessed with an automated bias field inhomogeneity correction (Arnold et al, 2001) and geometrically normalized using 3 anatomical landmarks (anterior commissure (AC), posterior commissure (PC), and mid-sagittal point). The scans were reoriented and resliced so that anatomical labeling could be done in coronal planes that follow the AC-PC axis. The manual outlining was performed using an in-house software called NVM and the exact specification of each region of interest is defined in 1) Neuromorphometrics' General Segmentation Protocol (http://neuromorphometrics.com/Seg/) ,and 2) the BrainCOLOR Cortical Parcellation Protocol (http://Neuromorphometrics.com/ParcellationProtocol_2010-04-05.PDF). It has to be noted that the exact protocols used to label the scans evolved over time. Because of this, not all anatomical regions were labeled in every group (label number range: max=142, min=136).

**2.2. Dataset correction**

Right after downloading the Neuromorphometrics dataset, we performed a rigorous quality control of the dataset. We discovered that this dataset presented several issues that had to be corrected before using it.

**2.2.1. Image resolution, orientation and size**

After checking each individual file, we found that they had different acquisition orientations (coronal, sagittal and axial). They also have different resolutions (1x1x1, 0.95x0.93x1.2, 1.26x1.24x12, etc.) and different volume sizes (256x256x307, 256x256x299, 256x256x160, etc.). To standardize them we affine registered all image and corresponding label files to the MNI152 space using ANTS software which resulted in a homogeneous dataset with axial orientation, 1x1x1 mm$^3$ voxel resolution and a volume size of 181x217x181 voxels. We also checked the image quality and we removed 14 cases from the original dataset that presented strong image artifacts and severe blurring effects. This resulted in a final dataset of 100 cases.

### 2.2.2. Inconsistent and different number of labels

The selected 100 files from the previous step had 129 common labels from a total of 142 labels. After analyzing these 13 inconsistent labels we decided to treat each of them in a specific manner according to the detected issue. Label file description assigns label numbers from 1 to 207. However, we found that labels 228, 229, 230 and 231 were present in some files. After checking them we realized that labels 228 and 229 corresponded to left a right basal foreground (labels 75 and 76) so we renumbered them. Labels 230 and 231 just represented few pixels in 3 of the cases and therefore were removed. Labels 63 and 64 (right and left vessel) were not present in all the cases (not always visible) and we decided to renumber them as part of the putamen (labels 57 and 58) as they were inside of it. We removed label 69 (optic chiasm) because it was not present in all the cases and its delineation was very inconsistent. Labels 71, 72 and 73 (cerebellar vermal lobules I-V, VI-VII and VIII-X) were present in 74 of the 100 cases and we decided to re-segment the inconsistent cases so all the cases have these labels (details will be given below). Label 78 (corpus callosum) was only present in 25 cases and we decided to relabel it as right and left white matter (labels 44 and 45). Label 15 ($5^{th}$ ventricle) was very tiny and only present in few cases (13), thus it was relabeled as lateral ventricles (labels 51 and 52). Finally, we decided to add two new labels that we found important. External CSF (labelled as 1) and left and right white matter lesions (labels 53 and 54). Details on how these labels were added are provided bellow. After all the cleanup the final dataset had a consistent number of 135 labels (see appendix).

### 2.2.3. Labelling errors

Once the dataset had a homogeneous number of labels, we inspected them to check their quality. After inspecting the dataset visually, we found that boundaries of all the structures in sagittal and axial planes were very irregular. This is probably due to the fact that the original manual delineation was performed in coronal plane. However, one of the main problems we found was the fact that cortical grey matter was severely overestimated and correspondently the CSF and white matter underestimated. This fact has been already highlighted by other researchers (Huo et al., 2017) that pointed out this problem in the context of cortical thickness estimation. The same problem arises in the cerebellum although is a bit less pronounced. To solve this problem (Huo et al., 2017) used an automatic fusion of the original GM/WM maps and partial volume maps generated by the method TOADS (Bazin and Pham, 2008) to correct the cortical labels. In this work we have followed a different approach based in the original manual segmentation and the intensity information.

First, we combined all the 135 labels into 7 different classes (cerebrospinal fluid (CSF), cortical grey matter (cGM), cerebral white matter (cWM), subcortical grey matter (sGM), cerebellar grey matter (ceGM), cerebellar white matter (ceWM) and brainstem (B)). External CSF was not labeled in Neuromorphometrics dataset so we added it using volBrain (Manjón et al., 2016) (we copied CSF label to those pixels that had label 0 in the original label file). Then, the median value of cGM and cWM was estimated and used to generate the partial volume maps using a linear mixing model (Manjón et al., 2008). Voxels in the cGM and cWM interface were relabeled according to their partial volume content (for example a cGM voxel with a cWM partial volume coefficient bigger than its corresponding cGM partial volume coefficient was relabeled as cWM). The same process was repited for the CSF/cGM interface, ceGM/ceWM interface and ceGM/CSF interface. To ensure the regularity of the new label maps each partial volume map was regularized using a non-local means filter (Coupe et al.,2018). Finally, each case was visually revised and small labelling errors were manually corrected using ITK-SNAP software. Most of the corrections were related with cortical grey matter in the upper part of the brain, misclassifications of white matter lesions as cortical grey matter and CSF related corrections. Figure 1 shows an example of the cGM/cWM tissue maps before and after the correction.

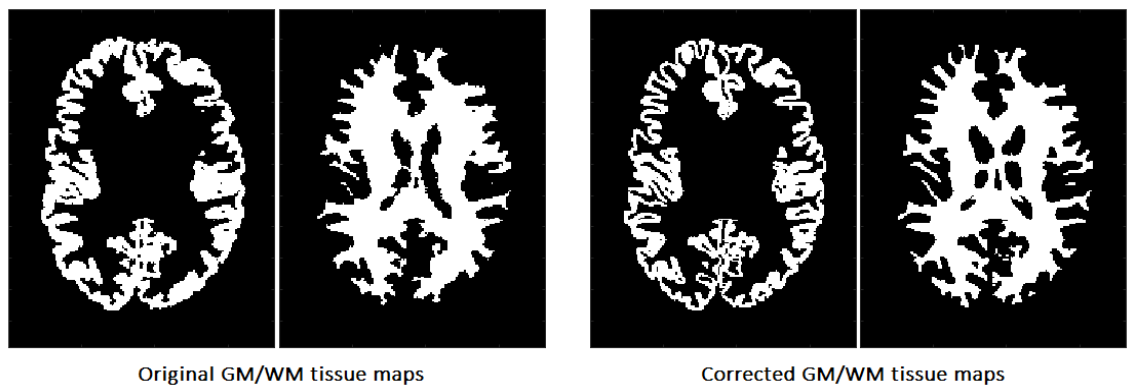

Original GM/WM tissue maps    Corrected GM/WM tissue maps

*Figure 1. Example of* cGM/cWM *tissue correction.*

After the tissue correction the original structure labels were automatically relabeled to match the new tissue maps. Specifically, those voxels that kept the same tissue type before and after the correction kept their original labels and those that changed were automatically labeled according the most likely label considering their position and intensity. Results were visually reviewed to assess its correctness and manually corrected when necessary. Finally, we realized that subcortical grey matter structures showed important segmentation errors and we decided to re-segment them using volBrain automatic segmentation followed by manual correction when needed. Figure 2 shows an example of the final relabeling result.

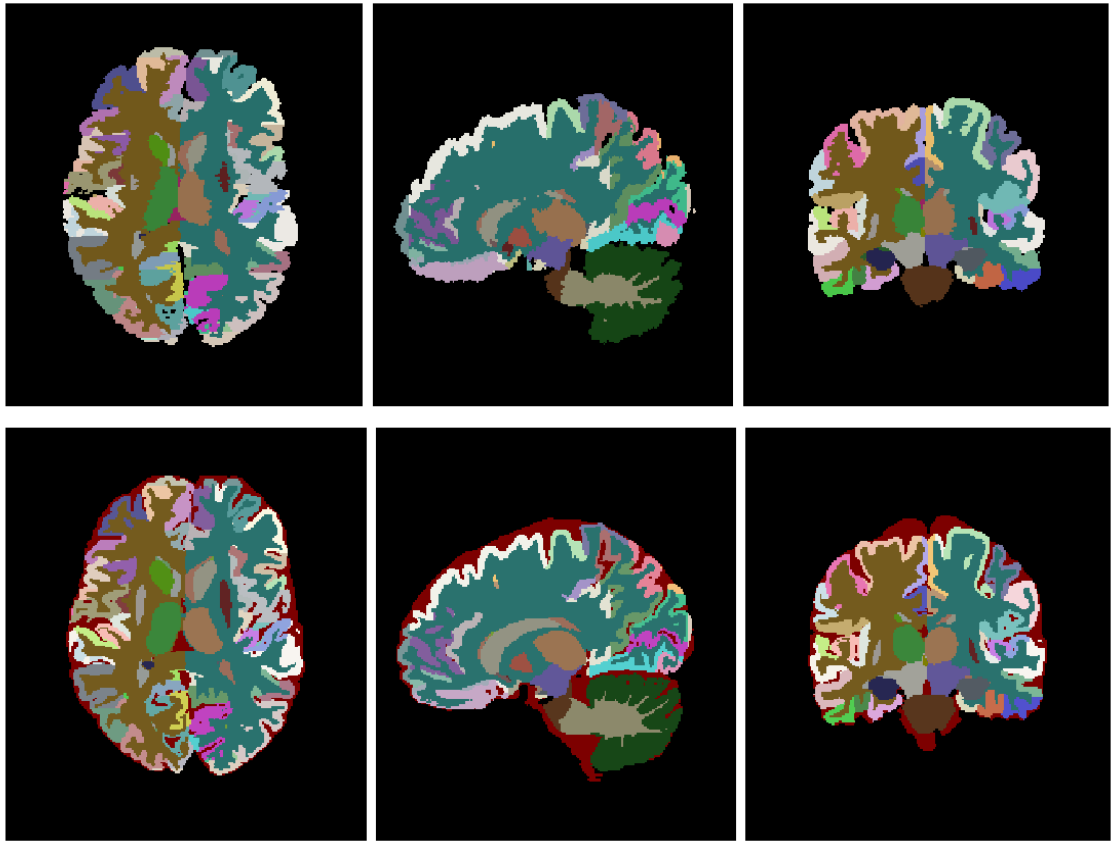

*Figure 2. Top row shows the original labelling and bottom row the corrected labeling. Note that external CSF label has been added to the labeling protocol.*

## 2.3. LesionBrain dataset

One of the main goals of the proposed pipeline was to make it robust to the presence of white matter (WM) lesions that normally are misclassified as GM in pathological brains. To this end, we included in our library not only healthy cases but also subjects with WM lesions. Specifically, 32 of the 100 cases of the previously described Neuromorphometrics dataset had WM visible lesions with a lesion load range from moderate to severe. We are aware that WM lesions can appear anywhere in the brain but it is also known that they have a priori probability to be located in periventricular areas among others (Coupe et al., 2018).

We found though that the number of cases with lesions on the dataset was not enough to capture the diversity of WM lesion distribution so we decided to expand the dataset using a manually labeled multiple sclerosis dataset. We previously used this dataset to develop a multiple sclerosis (MS) segmentation method (Coupe et al., 2018).

This dataset is composed of 43 MS patients who underwent 3T 3D-T1w MPRAGE and 3D-Fluid-Attenuated Inversion Recovery (FLAIR) MRI. We used only the T1 images since this is the input modality of our proposed pipeline. To further increase the dataset size, we included the left-right flipped version of the images and labels resulting in an extended dataset of 86 cases.

**2.4 vol2Brain pipeline description**

The vol2Brain pipeline is a set of image processing tasks dedicated to improve the quality of the input data and to set them into a specific geometric and intensity space, to segment the different structures and to generate useful volumetric information (see figure 3 for a general overview). The vol2Brain pipeline is based on the following steps:

1. Preprocessing
2. Multiscale labelling and cortical thickness estimation
3. Report and csv generation

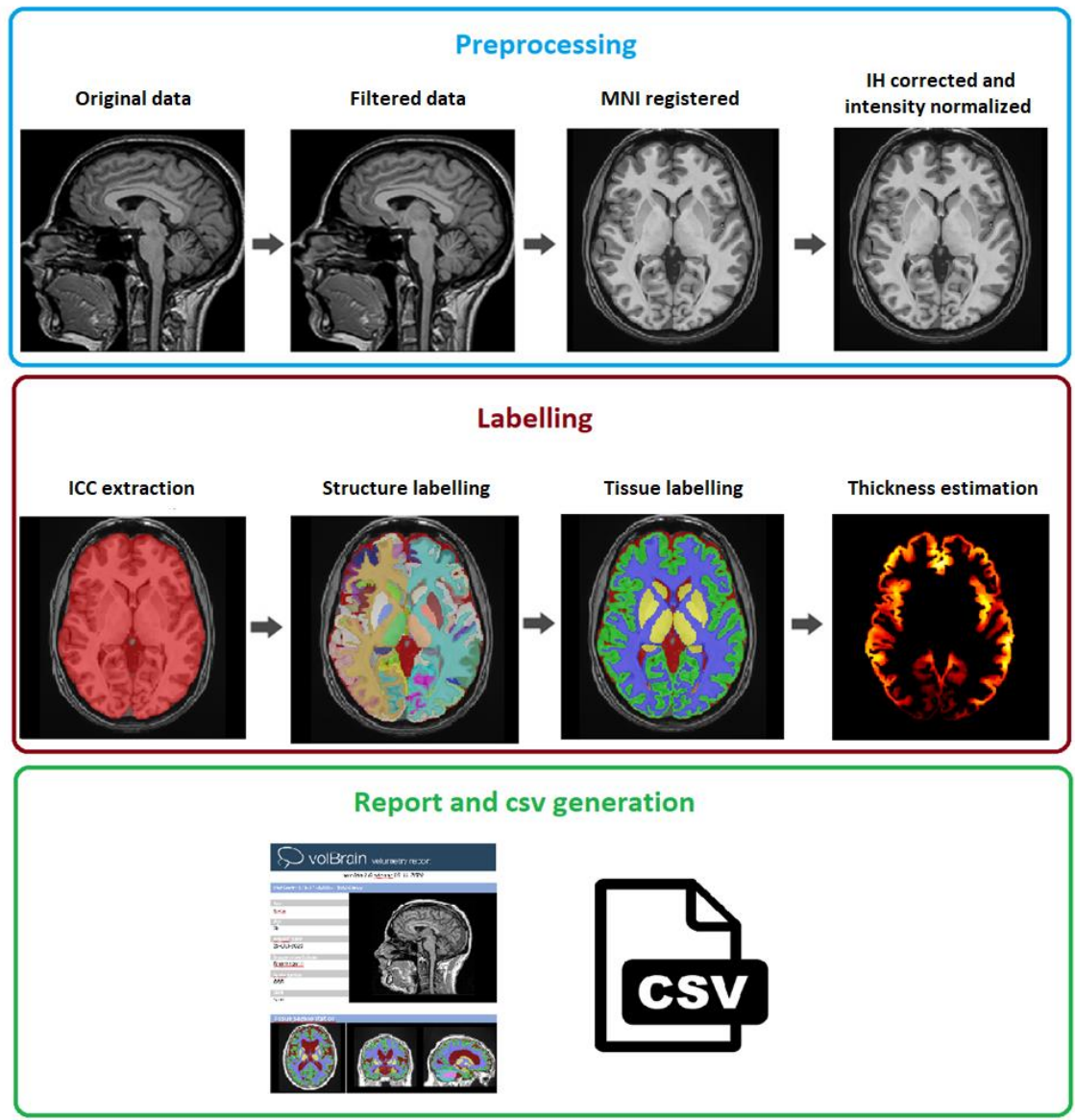

***Figure 3.*** *vol2Brain pipeline scheme. In the first row, the preprocessing for any new subject is presented. In the second row, the result of the ICC extraction, structure and tissue segmentations jointly with the cortical thickness estimation are presented. Finally, in the third row the volumetric information is extracted and presented.*

**2.3.1 Preprocessing**

We have used the same preprocessing steps than volBrain pipeline (Manjón and Coupe, 2016) since it has been demonstrated to be very robust (based in our experience processing more than 360.000 subjects worldwide). This preprocessing consists of the following steps. To improve the image quality, first the raw image is denoised using the Spatially Adaptive Non-Local Means (SANLM) filter (Manjón et al., 2010) and inhomogeneity corrected using the N4 method (Tustison et al., 2010). The resulting image is then affinely registered to the Montreal Neurological Institute (MNI) space using the ANTS software (Avants et al., 2008). The image in the MNI space have a size of 181x217x181 voxels with 1 mm$^3$ voxel resolution. Then we used a second inhomogeneity correction based on SPM8 (Ashburner and Friston, 2005) toolbox since this model-based method has proven to be quite robust once the data is located at the MNI space. Finally, we intensity normalized the images applying a piecewise linear tissue intensity mapping based on the TMS method (Manjón et al.,2008) as described in (Manjón and Coupe, 2016). It is worth to note that the library images were also intensity normalized using the described approach so both library and the case to be segmented share a common geometrical and intensity space.

**2.3.2. Multiscale labelling and cortical thickness estimation**

After the preprocessing, the images are ready to be segmented and measured. This segmentation is performed in several stages.

**ICC Extraction**

The first step in the segmentation process is the intracranial cavity extraction (ICC). This is obtained using the method NICE (Manjón et al., 2014). NICE method is based on a multi-scale non-local label fusion scheme. Details of NICE method can be found in Manjón et al. (2014). To further improve the quality of the original NICE method we have increased the size of the original volBrain template library from 100 to 300 cases using the 100 cases of the vol2Brain library and their left-right mirrored version.

**Full brain structure segmentation**

The dense segmentation of the full brain is based on a multiscale version the non-local patch-based label fusion technique (Coupe et al., 2011) where patches of the subject to be segmented are compared with patches of the training library to look for similar patterns within a predefined search volume to assign the proper label *v* as can be seen in equation 1.

$$v(x_i) = \frac{\sum_{s=1}^{N} \sum_{j \in V_i} w(x_i, x_{s,j}) y_{s,j}}{\sum_{s=1}^{N} \sum_{j \in V_i} w(x_i, x_{s,j})} \qquad (1)$$

where $V_i$ corresponds to the search volume, *N* is the number of subjects in the templates library, $y_{s,j}$ is the label of the voxel $x_{s,j}$ at the position *j* in the library subject *s* and $w(x_i, x_{s,j})$ is the patch similarity defined as:

$$w(x_i, x_{s,j}) = exp^{\frac{-D_{i,j,s}}{h^2}} \tag{2}$$

$$D_{i,j,s} = \left\lVert P(x_i) - P(x_{s,j}) \right\rVert_2^2 \tag{3}$$

where P(x_i) is the patch centered at *x_i*, *P(x_{s,j})* the patch centered at *x_j* in the templates and ||.||_2 is the normalized L2 norm (normalized by the number of elements) calculated from the distance between each pair of voxels from both patches *P(x_i)* and *P(x_{s,j})*. *h* is a normalization parameter that is estimated from the minimum of all patch distances within the search volume.

However, exhaustive patch comparison process is very time consuming (even in reduced neighborhoods, i.e., when the search volume V is small). To reduce the computational burden of this process, we have used a multiscale adaptation of the OPAL method (Giraud et al., 2016) previously proposed in (Romero et al., 2017) which takes benefit from the concept of Approximate Nearest Neighbor Fields (ANNF). To further speed up at the process we processed only those voxels that were segmented as ICC by the NICE method.

In patch-based segmentation, the patch size is a key parameter that is strongly related to the structure to be segmented and image resolution. It can be seen in the literature that multi-scale approaches improve segmentation results (Manjón et al., 2014). In OPAL (Giraud et al., 2016) independent and simultaneous multi-scale and multi-feature ANN fields were computed. Thus, we have followed a multi-scale approach were several different ANNs are computed for different patch-size resulting in different label probability maps that have to be combined. In this paper, two patch-size are used and an adaptive weighting scheme is proposed to fuse these maps (Eq. 3).

$$p(l) = \alpha \, p_1(l) + (1 - \alpha) p_2(l) \tag{3}$$

where *p_1(l)* is the probability map of patch-size 3x3x3 volxels for label *l*, *p_2(l)* the probability map of patch-size 5x5x5 voxels for label l, *p(l)* is the final probability map for label l and α ∈ [0,1] is the probability mixing coefficient.

**Systematic error correction**

Any segmentation method is subject to both random and systematic errors. The first error type can be typically minimized by using bootstrapped estimations. Fortunately, the non-local label fusion technique estimates the voxel label averaging the votes of many patches which naturally reduces the random classification error. Unfortunately, systematic errors cannot be reduced using this strategy since they are not random. However, due its nature this systematic bias can be learned and later use this knowledge to correct the segmentation output (Wang et al., 2013).

In (Romero et al,2017), we proposed an error corrector method based on a patch-based ensemble of neural networks (PEC for Patch-based Ensemble Corrector) to increase the segmentation accuracy by reducing the systematic errors. Specifically, a shallow neural network ensemble is trained with image patches of sizes 3x3x3 voxels (fully sampled) and 7x7x7 voxels (subsampled by skipping two voxels at each dimension) from the T1w images, the automatic segmentations, a distance map value, and their x, y and z coordinates at MNI152 space. The distance map we used is calculated for the whole structure as the distance in voxels to the structure contour. This results in a feature vector of 112 features that are mapped to a patch of manual segmentations of size 3x3x3 voxels. We used a multilayer perceptron with two hidden layers of size 83 and 55 neurons resulting in a network with a topology of 112x83x55x27 neurons. An ensemble of 10 neural networks was trained using a boosting strategy. Each new network was trained with a different subset of data which was selected by giving a higher probability of selection to those samples that were misclassified in the previous ensemble. More details can be found in the original paper (Romero et al,2017).

**2.3.3. Multiscale label generation**

Once the full brain segmentation is performed different scale versions were computed combining several labels to generate more generic ones and allow a multiscale brain analysis. The 135 labels were combined to create a tissue type segmentation map including 8 different tissues (cerebrospinal fluid (CSF), cortical grey matter (cGM), cerebral white matter (cWM), subcortical grey matter (sGM), cerebellar grey matter (ceGM), cerebellar white matter (ceWM), brainstem (BS) and white matter lesions (WML)). The cGM and cWM maps will be later used to compute the cortical thickness. Also, cerebrum lobe maps were created by combining cortical GM structures. These maps will be used later to computed lobe specific volumes and thickness.

**2.3.4. Cortical thickness estimation**

To estimate the cortical gray matter thickness, we have used the DiReCT method. DiReCT was introduced in Das et al. (2009) and was made available in ANTs under the program named *KellyKapowski*. This method is based on the use of a dense nonlinear registration to estimate the distance between the inner and the outer parts of the cortical gray matter. Cortical thickness per cortical label and per lobe were estimated from the thickness map and the corresponding segmentation maps (Tustison et al.,2014).

### 2.3.5. Report generation

The output produced by the vol2Brain pipeline consists in a pdf and csv files. These files summarize the volumes and asymmetry ratios estimated from the images. If the user provides sex and age of the submitted subject population-based normal volumes and asymmetry bounds for all structures are added for reference purposes. This normality bounds were automatically estimated from the IXI dataset (https://brain-development.org/ixi-dataset/) which contains almost 600 normal subjects covering most of the adult lifespan.

Furthermore, the user can access to its user area through volBrain website to download the resulting Nifty files containing the segmentations at different scales (both in native or MNI space). An example of the volumetric report produced by vol2Brain is shown in the appendix.

## 3. Experiments and results

In this section some experimental results are shown to highlight the accuracy and reproducibility of the proposed pipeline. A leave-two-out procedure was performed for the 100 subjects of the library (i.e. excluding the case to be segmented and its mirrored version). In the dataset there are 19 cases that were scanned and labelled twice for reproducibility estimation purposes. In this case, a leave-four-out procedure was applied to avoid any problem (i.e. excluding the case to be segmented and its mirrored version of the two acquisitions of the same subject). To measure the segmentation quality the dice index (Zijdenbos et al., 1994) was computed by comparing the manual segmentations with the segmentations obtained with our method. A visual example of the automatic segmentation results is shown in Figure 4.

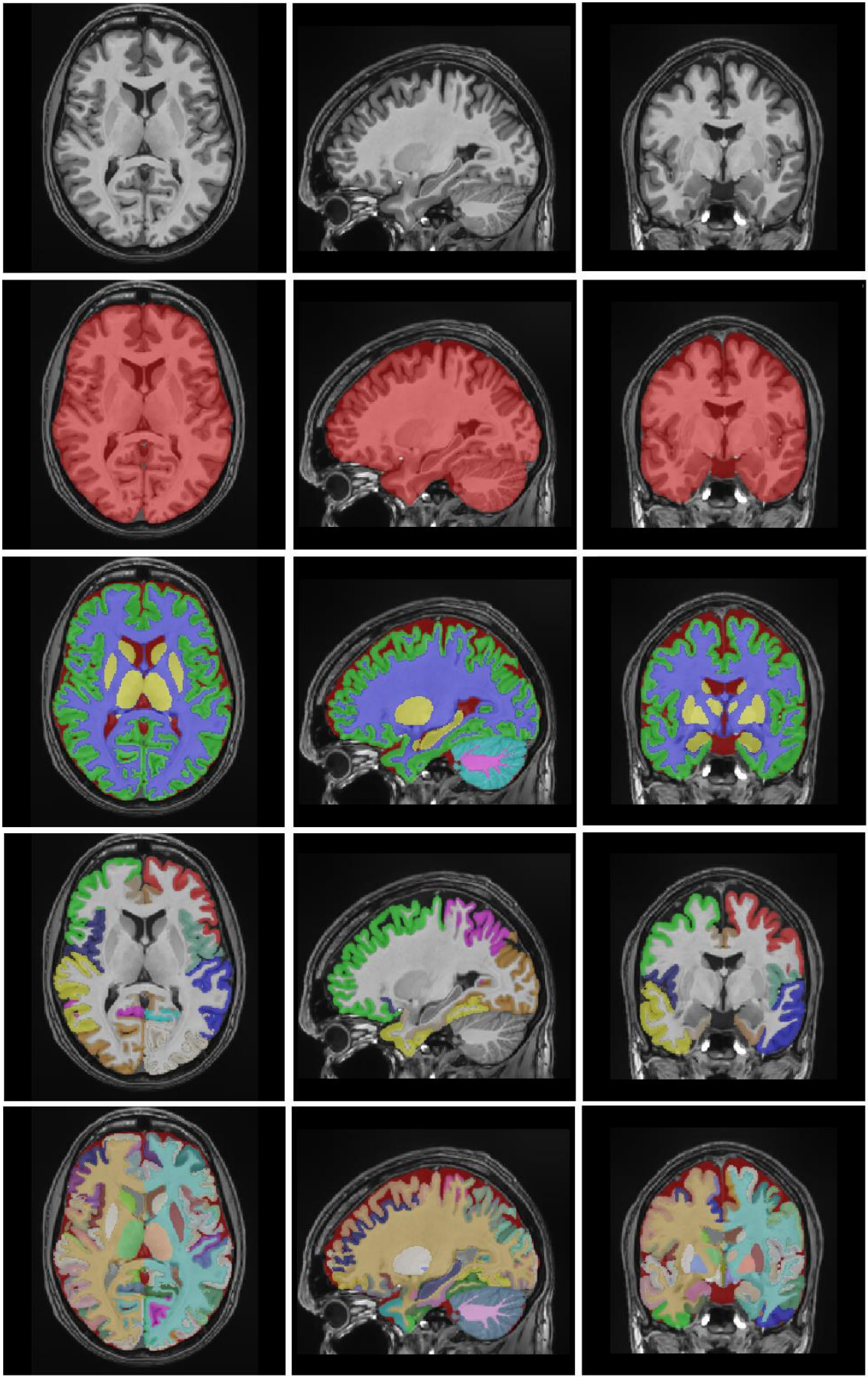

***Figure 4.** Example results of vol2Brain. T1 image, ICC mask, brain tissues, lobes and structures.*

## 3.1. Results

Since presenting dice results of the 135 labels would be impractical, we have decided to show the average results for cortical and non-cortical labels as done in previous papers (Wang and Yushkevich, 2013). In Table 1, the results of the proposed method are shown with and without the corrective learning step (PEC) to show the impact that this postprocessing has in the final results (it improved the results in all the cases).

Table 1: Proposed method dice results. The mean Dice is evaluated on the all considered labels (135 without background).

| Method | All labels | Cortical labels | Noncortical labels |
|---|---|---|---|
| Our method | 0.8190 ± 0.0300 | 0.7912 ± 0.0397 | 0.8929 ± 0.0173 |
| Our method + PEC | **0.8262 ± 0.0257** | **0.7996 ± 0.0347** | **0.8969 ± 0.0157** |

To further explore the results, we separated them by dataset as it is well known that results within the same dataset are normally better than across datasets. This allows to explore the generalization capabilities of the proposed method. Results are summarized in Table 2. As can be seen, results of the OASIS dataset were the best among datasets. This makes perfect sense since precisely this dataset is the biggest. CANDI dataset showed the worst results. This dataset had the worst image quality which somehow explains these results.

Table 2: Proposed method overall dice results for the full dataset and for each of the subsets.

| All (N=100) | OASIS (N=68) | CANDI (N=6) | ADNI (N=25) | COLIN (N=1) |
|---|---|---|---|---|
| 0.8262 ± 0.0257 | 0.8353± 0.0233 | 0.7831± 0.0326 | 0.8111 ± 0.0142 | 0.8353 |

One of the objectives of the proposed method was to be able to deal with images with white mater lesions. This is fundamental since if we do not take into account those regions, they are normally misclassified a cortical or subcortical grey matter (and also affects the cortical thickness estimation) (Dadar et al., 2021). In Table 3 the results of white matter lesion segmentation are shown (left and right lesions were considered together). We separated the results by lesion volume as it is well known that small lesions are more difficult to segment than big ones (Manjón et al., 2018).

Table 3: Proposed method lesion dice results. *Small(<4 ml), Medium(4 ml to 18 ml), Big(>18 ml).

| Method | Small (N=76) | Medium (N=21) | Big (N=3) | Avg (N=100) |
|---|---|---|---|---|
| lesion | 0.5767± 0.1486 | 0.8281± 0.0500 | 0.8467± 0.0524 | 0.6440 ± 0.1589 |

Once the method segments the full brain into 135 labels, those labels are grouped together to provide information at different anatomical scales. Specifically, eight different tissue labels are generated. Dice results are shown in Table 4.

Table 4: Proposed method dice results for each brain tissue.

| CSF | cGM | cWM | sGM |
|---|---|---|---|
| 0.9006 ± 0.0307 | 0.9543 ± 0.0144 | 0.9669 ± 0.0131 | 0.9518 ± 0.0114 |
| **ceGM** | **ceWM** | **BS** | **Lesion** |
| 0.9644 ± 0.0172 | 0.9448 ± 0.0363 | 0.9693 ± 0.0137 | 0.6440 ± 0.1589 |

### 3.2. Method reproducibility

A very important feature for a measurement method is its reproducibility. To measure the reproducibility of the proposed method, we used a subset of our library. Specifically, we used 19 cases of the OASIS subset that were scanned and labeled twice. In this case we have two sources of variability which are related to the inter-image changes and manual labelling differences. To measure the reproducibility, we computed the dice coefficient between the two different segmentations (of each case and its repetition). This was done for both, the manual segmentation (that we used as reference) and the automatic one. Results are summarized in Table 5. As can be seen the proposed method showed a slightly superior reproducibility than manual labeling.

Table 5: Proposed method dice results. The mean Dice is evaluated on the all considered labels (135 without background).

| Method | All labels | Cortical labels | Noncortical labels |
|---|---|---|---|
| **vol2Brain** | 0.8405 ± 0.0181 | 0.8234 ± 0.0206 | 0.8856 ± 0.0158 |
| **Manual** | 0.8368 ± 0.0171 | 0.8198 ± 0.0200 | 0.8818 ± 0.0163 |

### 3.3. Method comparison

It is difficult to compare the proposed method with similar state of the art methods such as Freesurfer since the labelling protocol is slightly different. For this reason, we have used as reference a freely available and well-known method call Joint Label Fusion (Wang and Yushkevich, 2013). This method is a state-of-the-art multi-atlas segmentation approach. To make it fully comparable we used as atlas library the corrected cases of our library. In Table 6, we show the results of the comparison. We compared our proposed method with two versions of the JLF approach, one using an affine registered library (linear) and another using a non-linear registered library. It is worth to note the proposed method uses only linearly registered library (i.e. no non-linear registration was used). As can be noticed the proposed method was far superior to both versions.

Table 6: Proposed method dice results compared to the results of two versions of JLF method.

| Method | All labels | Cortical labels | Noncortical labels |
|---|---|---|---|
| vol2Brain | 0.8262 ± 0.0257 | 0.7996 ± 0.0347 | 0.8969 ± 0.0157 |
| JLF (linear) | 0.7369 ± 0.0292 | 0.7016 ± 0.0337 | 0.8305 ± 0.0241 |
| JLF (non-linear) | 0.7591 ± 0.0252 | 0.7327± 0.0288 | 0.8291 ± 0.0228 |

### 3.4. Computational time

The proposed method takes around 20 minutes in average to complete the whole pipeline (including cortical thickness estimation and report generation). JLF method takes around 2 hours only for structure segmentations without cortical thickness estimation (excluding the preprocessing which includes several hours of non-linear registration depending on the number of atlases used). Freesurfer normally takes around 6 hours to perform the complete analysis (which also includes surface extraction).

## 4. Discussion

We have presented a new pipeline for full brain segmentation (vol2Brain) that is able to segment the brain into 135 different regions in a very efficient and accurate manner. The proposed method also integrates these 135 regions to provide measures at different anatomical scales, including brain tissues and lobes. It also provides cortical thickness measurements per cortical structure and lobe displayed into an automatic report summarizing the results (see appendix).

To create vol2Brain pipeline we had to create a template library that integrates all the anatomical information needed to perform the labelling process. This was a long and laborious work since the original library obtained from Neuromorphometrics did not meet the required quality and we had to invest a significant amount of time to make it ready to use. To create this library, we homogenized the image resolution, orientation and size of the images, removed and relabeled inconsistent labels and corrected systematic labelling errors. Besides, we extended the labelling protocol adding external CSF and white matter lesions. As a result, we generated a highly consistent and high-quality library that not only allowed to develop the current proposed pipeline but will be also a valuable resource for future developments.

The proposed method is based on patch-based multi-atlas label fusion technology. Specifically, we have used an optimized version of non-local label fusion called OPAL that efficiently finds patch matches needed to label each voxel in the brain reducing the required time to label the full brain from hours to minutes. To further improve the results, we have used a patch-based error corrector which has been previously used in other segmentation problems such as hippocampus subfield labelling (Romero et al., 2017) or cerebellum lobules (Carass et al, 2018).

We measured the results of the proposed pipeline using a LOO methodology and got an average dice value of 0.8262. This result was obtained from 4 different sub datasets ranking from 0.7831 to 0.8353 showing a good generalization of the proposed method. This result was quite close to the manual intra-observer accuracy that was estimated in 0.8363 using a reduced dataset. We also compared the proposed method with a related currently available state-of-the-art method for full brain labelling. We demonstrated that vol2Brain was not only far superior to the linear (0.8262 vs 0.7369) and non-linear (0.8262 vs 0.7591) versions of JLF method but also more efficient with a temporal cost of minutes compared to hours.

The proposed vol2Brain pipeline is already available though our volBrain platform (https://volbrain.upv.es). As the rest of the volBrain platform pipelines it receives an anonymized and compressed nifty file thought the website and reports the results 20 minutes later by sending an email to the user. The user can also download the segmentation files though the user area of volBrain platform (an example of the pdf report is shown in the appendix).

We hope that the accuracy and efficiency of the proposed method and the ease of use though the volBrain platform will boost the anatomical analysis of the normal and pathological brain (specially on those cases with white matter lesions).

## 5. Conclusion

In this paper, we present a novel pipeline to densely segment the brain and to provide measurements of different features at different anatomical scales in an accurate and efficient manner. The proposed pipeline has been compared with a state-of-the-art related method showing competitive results in term of accuracy and computational time. Finally, we hope that the online accessibility of the proposed pipeline will facilitate the access of any user around the world to the proposed pipeline making their MRI data analysis simpler and more efficient.

# Acknowledgments

This research was supported by the Spanish DPI2017-87743-R grant from the Ministerio de Economia, Industria y Competitividad of Spain. This work benefited from the support of the project DeepvolBrain of the French National Research Agency (ANR-18-CE45-0013). This study was achieved within the context of the Laboratory of Excellence TRAIL ANR-10-LABX-57 for the BigDataBrain project. Moreover, we thank the Investments for the future Program IdEx Bordeaux (ANR-10- IDEX- 03- 02, HL-MRI Project), Cluster of excellence CPU and the CNRS.

# Appendix

In table 1 a list of the 135 labels measured in vol2Brain pipeline is shown.

*Table 1. List of the structures labelled in vol2Brain pipeline.*

| Index | Label Number | Label Name |
|---|---|---|
| 1 | 1 | CSF |
| 2 | 4 | 3rd Ventricle |
| 3 | 11 | 4th Ventricle |
| 4 | 23 | Right Accumbens |
| 5 | 30 | Left Accumbens |
| 6 | 31 | Right Amygdala |
| 7 | 32 | Left Amygdala |
| 8 | 35 | Brain Stem |
| 9 | 36 | Right Caudate |
| 10 | 37 | Left Caudate |
| 11 | 38 | Right Cerebellum Exterior |
| 12 | 39 | Left Cerebellum Exterior |
| 13 | 40 | Right Cerebellum White Matter |
| 14 | 41 | Left Cerebellum White Matter |
| 15 | 44 | Right Cerebral White Matter |
| 16 | 45 | Left Cerebral White Matter |
| 17 | 47 | Right Hippocampus |
| 18 | 48 | Left Hippocampus |
| 19 | 49 | Inferior Right Lateral Ventricle |
| 20 | 50 | Inferior Left Lateral Ventricle |
| 21 | 51 | Right Lateral Ventricle |
| 22 | 52 | Left Lateral Ventricle |
| 23 | 53 | Right Lesion |
| 24 | 54 | Left Lesion |
| 25 | 55 | Right Pallidum |
| 26 | 56 | Left Pallidum |
| 27 | 57 | Right Putamen |
| 28 | 58 | Left Putamen |
| 29 | 59 | Right Thalamus |
| 30 | 60 | Left Thalamus |
| 31 | 61 | Right Ventral DC |
| 32 | 62 | Left Ventral DC |
| 33 | 71 | Cerebellar Vermal Lobules I-V |
| 34 | 72 | Cerebellar Vermal Lobules VI-VII |
| 35 | 73 | Cerebellar Vermal Lobules VIII-X |
| 36 | 75 | Left Basal Forebrain |
| 37 | 76 | Right Basal Forebrain |
| 38 | 100 | Right anterior cingulate gyrus |
| 39 | 101 | Left anterior cingulate gyrus |
| 40 | 102 | Right anterior insula |
| 41 | 103 | Left anterior insula |
| 42 | 104 | Right anterior orbital gyrus |
| 43 | 105 | Left anterior orbital gyrus |
| 44 | 106 | Right angular gyrus |
| 45 | 107 | Left angular gyrus |
| 46 | 108 | Right calcarine cortex |
| 47 | 109 | Left calcarine cortex |
| 48 | 112 | Right central operculum |
| 49 | 113 | Left central operculum |
| 50 | 114 | Right cuneus |
| 51 | 115 | Left cuneus |
| 52 | 116 | Right entorhinal area |
| 53 | 117 | Left entorhinal area |
| 54 | 118 | Right frontal operculum |
| 55 | 119 | Left frontal operculum |
| 56 | 120 | Right frontal pole |
| 57 | 121 | Left frontal pole |
| 58 | 122 | Right fusiform gyrus |
| 59 | 123 | Left fusiform gyrus |
| 60 | 124 | Right gyrus rectus |
| 61 | 125 | Left gyrus rectus |
| 62 | 128 | Right inferior occipital gyrus |
| 63 | 129 | Left inferior occipital gyrus |
| 64 | 132 | Right inferior temporal gyrus |
| 65 | 133 | Left inferior temporal gyrus |
| 66 | 134 | Right lingual gyrus |
| 67 | 135 | Left lingual gyrus |
| 68 | 136 | Right lateral orbital gyrus |
| 69 | 137 | Left lateral orbital gyrus |
| 70 | 138 | Right middle cingulate gyrus |
| 71 | 139 | Left middle cingulate gyrus |
| 72 | 140 | Right medial frontal cortex |
| 73 | 141 | Left medial frontal cortex |
| 74 | 142 | Right  middle frontal gyrus |
| 75 | 143 | Left middle frontal gyrus |
| 76 | 144 | Right middle occipital gyrus |
| 77 | 145 | Left middle occipital gyrus |
| 78 | 146 | Right medial orbital gyrus |
| 79 | 147 | Left medial orbital gyrus |
| 80 | 148 | Right postcentral gyrus medial segment |
| 81 | 149 | Left postcentral gyrus medial segment |
| 82 | 150 | Right precentral gyrus medial segment |
| 83 | 151 | Left precentral gyrus medial segment |
| 84 | 152 | Right superior frontal gyrus medial segment |
| 85 | 153 | Left superior frontal gyrus medial segment |
| 86 | 154 | Right middle temporal gyrus |
| 87 | 155 | Left middle temporal gyrus |
| 88 | 156 | Right occipital pole |
| 89 | 157 | Left occipital pole |
| 90 | 160 | Right occipital fusiform gyrus |
| 91 | 161 | Left occipital fusiform gyrus |
| 92 | 162 | Right opercular part of the inferior frontal gyrus |
| 93 | 163 | Left opercular part of the inferior frontal gyrus |
| 94 | 164 | Right orbital part of the inferior frontal gyrus |
| 95 | 165 | Left orbital part of the inferior frontal gyrus |
| 96 | 166 | Right posterior cingulate gyrus |
| 97 | 167 | Left posterior cingulate gyrus |
| 98 | 168 | Right precuneus |
| 99 | 169 | Left precuneus |
| 100 | 170 | Right parahippocampal gyrus |
| 101 | 171 | Left parahippocampal gyrus |
| 102 | 172 | Right posterior insula |
| 103 | 173 | Left posterior insula |
| 104 | 174 | Right parietal operculum |
| 105 | 175 | Left parietal operculum |
| 106 | 176 | Right postcentral gyrus |
| 107 | 177 | Left postcentral gyrus |
| 108 | 178 | Right posterior orbital gyrus |
| 109 | 179 | Left posterior orbital gyrus |
| 110 | 180 | Right planum polare |
| 111 | 181 | Left planum polare |
| 112 | 182 | Right precentral gyrus |
| 113 | 183 | Left precentral gyrus |
| 114 | 184 | Right planum temporale |
| 115 | 185 | Left planum temporale |
| 116 | 186 | Right subcallosal area |
| 117 | 187 | Left subcallosal area |
| 118 | 190 | Right superior frontal gyrus |
| 119 | 191 | Left superior frontal gyrus |
| 120 | 192 | Right supplementary motor cortex |
| 121 | 193 | Left supplementary motor cortex |
| 122 | 194 | Right supramarginal gyrus |
| 123 | 195 | Left supramarginal gyrus |
| 124 | 196 | Right superior occipital gyrus |
| 125 | 197 | Left superior occipital gyrus |
| 126 | 198 | Right superior parietal lobule |
| 127 | 199 | Left superior parietal lobule |
| 128 | 200 | Right superior temporal gyrus |
| 129 | 201 | Left superior temporal gyrus |
| 130 | 202 | Right temporal pole |
| 131 | 203 | Left temporal pole |
| 132 | 204 | Right triangular part of the inferior frontal gyrus |
| 133 | 205 | Left triangular part of the inferior frontal gyrus |
| 134 | 206 | Right transverse temporal gyrus |
| 135 | 207 | Left transverse temporal gyrus |

## Report example

Example report of the vol2Brain pipeline. The report summarizes volumetric/asymmetry and thickness information at different scales as long as different captures of the results as a visual quality control.

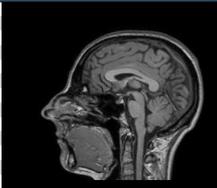
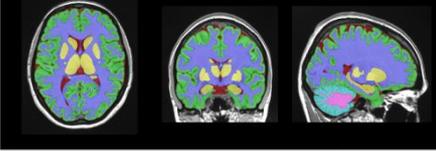
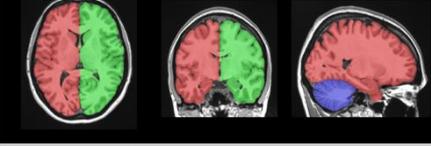
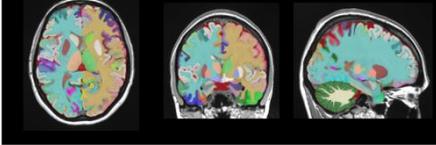

| CSF | Total (cm³/%) | Right (cm³/%) | Left (cm³/%) | Asymmetry (%) |
|---|---|---|---|---|
| Inf. Lateral Ventricle | 0.31 (0.023) [0.018, 0.026] | 0.20 (0.015) [0.000, 0.044] | 0.11 (0.009) [0.000, 0.035] | 54.1966 [-85.696, 120.7] |
| Lateral Ventricle | 7.88 (0.590) [0.000, 2.217] | 3.08 (0.231) [0.000, 1.124] | 4.80 (0.359) [0.000, 1.148] | -43.6402 [-59.556, 47.042] |
| 3rd Ventricle | 0.74 (0.055) [0.000, 0.112] | | | |
| 4th Ventricle | 1.58 (0.118) [0.044, 0.196] | | | |
| External CSF | 78.67 (5.888) [4.751, 13.032] | | | |

| Cerebellar vermis | Total (cm³/%) |
|---|---|
| Lobules I-V | 3.67 (0.274) [0.221, 0.374] |
| Lobules VI-VII | 1.92 (0.144) [0.109, 0.187] |
| Lobules VIII-X | 2.89 (0.217) [0.158, 0.250] |

## Cortical thickness

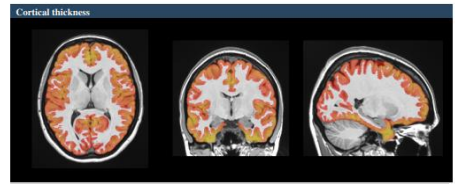

| Thickness | Total (mm/norm.) | Right (mm/norm.) | Left (mm/norm.) | Asymmetry (%) |
|---|---|---|---|---|
| **Frontal lobe** | 2.40 (0.022) [0.018, 0.026] | 2.33 (0.021) [0.018, 0.026] | 2.46 (0.022) [0.018, 0.026] | -5.5160 [-7.855, 4.645] |
| Frontal pole | 2.70 (0.025) [0.018, 0.028] | 2.80 (0.025) [0.018, 0.029] | 2.61 (0.024) [0.017, 0.028] | 7.2154 [-15.591, 24.153] |
| Gyrus rectus | 2.31 (0.021) [0.018, 0.056] | 2.25 (0.020) [0.019, 0.037] | 2.37 (0.022) [0.018, 0.056] | -5.0966 [-21.568, 31.650] |
| Opercular inf. frontal gyrus | 2.24 (0.020) [0.016, 0.024] | 2.17 (0.020) [0.015, 0.025] | 2.29 (0.021) [0.016, 0.024] | -5.5815 [-22.176, 25.578] |
| Orbital inf. frontal gyrus | 2.33 (0.021) [0.017, 0.026] | 2.47 (0.022) [0.017, 0.028] | 2.16 (0.020) [0.016, 0.026] | 13.1979 [-17.607, 30.989] |
| Triangular inf. frontal gyrus | 2.31 (0.021) [0.017, 0.025] | 2.39 (0.022) [0.017, 0.026] | 2.22 (0.020) [0.016, 0.024] | 7.3733 [-16.265, 27.125] |
| Medial frontal cortex | 2.50 (0.023) [0.022, 0.033] | 2.60 (0.024) [0.022, 0.034] | 2.36 (0.021) [0.021, 0.033] | 9.8348 [-15.753, 25.973] |
| Middle frontal gyrus | 2.60 (0.024) [0.018, 0.028] | 2.74 (0.025) [0.018, 0.028] | 2.46 (0.022) [0.018, 0.028] | 10.6628 [-12.286, 15.284] |
| Anterior orbital gyrus | 2.85 (0.026) [0.021, 0.033] | 2.90 (0.026) [0.020, 0.033] | 2.77 (0.025) [0.021, 0.034] | 4.5097 [-25.109, 16.660] |
| Lateral orbital gyrus | 2.70 (0.025) [0.020, 0.031] | 2.83 (0.026) [0.020, 0.032] | 2.59 (0.023) [0.019, 0.031] | 8.8353 [-16.021, 24.074] |
| Medial orbital gyrus | 2.15 (0.020) [0.018, 0.031] | 2.05 (0.019) [0.017, 0.032] | 2.25 (0.020) [0.019, 0.052] | -9.1979 [-22.493, 13.397] |
| Posterior orbital gyrus | 2.89 (0.026) [0.022, 0.035] | 2.97 (0.027) [0.022, 0.035] | 2.80 (0.025) [0.021, 0.035] | 6.0607 [-14.948, 18.509] |
| Precentral gyrus | 1.78 (0.016) [0.013, 0.019] | 1.79 (0.016) [0.013, 0.019] | 1.77 (0.016) [0.013, 0.020] | 1.2456 [-15.658, 12.829] |
| Precentral gyrus medial segment | 1.93 (0.018) [0.012, 0.021] | 1.95 (0.018) [0.012, 0.021] | 1.91 (0.017) [0.012, 0.021] | 1.7518 [-21.417, 22.271] |
| Subcallosal area | 1.52 (0.014) [0.011, 0.030] | 1.34 (0.012) [0.009, 0.029] | 1.69 (0.015) [0.012, 0.031] | -22.8917 [-42.975, 19.435] |

| **Parietal lobe** | 110.52 (8.271) [7.440, 9.118] | 54.63 (4.089) [3.690, 4.556] | 55.89 (4.182) [3.714, 4.597] | -2.2689 [-7.037, 5.451] |
|---|---|---|---|---|
| Angular gyrus | 15.81 (1.183) [1.184, 1.863] | 8.04 (0.601) [0.614, 0.990] | 7.78 (0.582) [0.540, 0.905] | 3.2855 [-10.876, 32.046] |
| Postcentral gyrus | 21.17 (1.584) [1.192, 1.635] | 10.19 (0.763) [0.560, 0.825] | 10.97 (0.821) [0.596, 0.846] | -7.3629 [-23.442, 15.193] |
| Postcentral gyrus medial segment | 2.00 (0.149) [0.083, 0.367] | 0.88 (0.066) [0.036, 0.089] | 1.11 (0.083) [0.036, 0.089] | -22.9557 [-55.494, 54.055] |
| Precuneus | 19.63 (1.469) [1.400, 1.925] | 10.18 (0.762) [0.683, 0.975] | 9.45 (0.707) [0.696, 0.971] | 7.4571 [-14.103, 12.920] |
| Sup. parietal lobule | 19.84 (1.485) [1.200, 1.683] | 9.98 (0.747) [0.568, 0.845] | 9.86 (0.738) [0.598, 0.873] | 1.2075 [-23.402, 15.431] |
| Supramarginal gyrus | 14.82 (1.109) [1.028, 1.447] | 7.21 (0.540) [0.498, 0.741] | 7.61 (0.570) [0.492, 0.745] | -5.4343 [-22.898, 23.638] |
| **Temporal lobe** | 93.26 (6.980) [6.600, 8.297] | 46.78 (3.501) [3.277, 4.108] | 46.48 (3.479) [3.291, 4.131] | 0.6421 [-7.102, 6.081] |
| Fusiform gyrus | 14.83 (1.110) [0.952, 1.407] | 8.39 (0.628) [0.460, 0.721] | 6.44 (0.482) [0.463, 0.726] | 26.3081 [-20.110, 20.933] |
| Planum polare | 4.00 (0.300) [0.236, 0.348] | 1.67 (0.125) [0.109, 0.176] | 2.33 (0.174) [0.115, 0.185] | -32.6851 [-33.023, 22.187] |
| Planum temporale | 2.65 (0.199) [0.220, 0.305] | 1.43 (0.107) [0.093, 0.188] | 1.22 (0.091) [0.108, 0.216] | 15.7405 [-55.254, 27.669] |
| Inf. temporal gyrus | 23.74 (1.777) [1.403, 1.999] | 11.91 (0.891) [0.672, 1.002] | 11.83 (0.885) [0.690, 1.039] | 0.6558 [-22.932, 16.303] |
| Middle temporal gyrus | 26.26 (1.966) [1.861, 2.443] | 12.74 (0.954) [0.929, 1.264] | 13.52 (1.012) [0.886, 1.227] | -5.9110 [-13.179, 20.828] |
| Sup. temporal gyrus | 16.05 (1.201) [0.874, 1.242] | 8.88 (0.664) [0.477, 0.656] | 7.18 (0.537) [0.399, 0.624] | 21.1964 [-18.135, 31.777] |
| Transverse temporal gyrus | 2.26 (0.169) [0.167, 0.310] | 0.90 (0.068) [0.072, 0.151] | 1.35 (0.101) [0.083, 0.171] | -39.9204 [-50.179, 24.720] |
| Temporal pole | 20.71 (1.550) [1.073, 1.631] | 9.95 (0.745) [0.550, 0.830] | 10.76 (0.805) [0.503, 0.821] | -7.7699 [-12.518, 21.458] |

| **Occipital lobe** | 65.91 (4.933) [4.493, 5.944] | 33.59 (2.514) [2.241, 3.028] | 32.32 (2.419) [2.189, 2.980] | 3.8703 [-10.077, 14.006] |
|---|---|---|---|---|
| Calcarine cortex | 5.57 (0.417) [0.283, 0.586] | 2.65 (0.198) [0.137, 0.297] | 2.92 (0.219) [0.136, 0.299] | -9.8334 [-26.836, 26.832] |
| Cuneus | 6.59 (0.493) [0.483, 0.765] | 2.99 (0.224) [0.227, 0.391] | 3.60 (0.270) [0.231, 0.399] | -18.6691 [-30.603, 26.967] |
| Lingual gyrus | 15.83 (1.185) [0.999, 1.408] | 7.77 (0.582) [0.487, 0.712] | 8.06 (0.603) [0.484, 0.724] | -3.6401 [-19.124, 19.086] |
| Occipital fusiform gyrus | 5.94 (0.444) [0.345, 0.613] | 3.43 (0.257) [0.163, 0.318] | 2.51 (0.188) [0.158, 0.319] | 31.1066 [-34.292, 35.331] |
| Inf. occipital gyrus | 12.05 (0.902) [0.711, 1.065] | 5.54 (0.415) [0.333, 0.553] | 6.51 (0.487) [0.343, 0.553] | -16.1014 [-28.413, 25.107] |
| Middle occipital gyrus | 8.21 (0.614) [0.552, 0.921] | 4.15 (0.311) [0.252, 0.446] | 4.05 (0.303) [0.275, 0.502] | 2.3718 [-38.852, 17.175] |
| Sup. occipital gyrus | 7.53 (0.564) [0.391, 0.662] | 3.75 (0.280) [0.201, 0.359] | 3.79 (0.283) [0.166, 0.326] | -1.0134 [-18.190, 45.435] |
| Occipital pole | 4.19 (0.313) [0.194, 0.460] | 2.04 (0.153) [0.080, 0.220] | 2.15 (0.161) [0.096, 0.259] | -5.2914 [-61.089, 26.958] |
| **Limbic cortex** | 37.60 (2.814) [2.678, 3.482] | 18.97 (1.420) [1.342, 1.777] | 18.62 (1.394) [1.289, 1.752] | 1.8516 [-11.178, 16.373] |
| Entorhinal area | 3.72 (0.279) [0.230, 0.354] | 1.77 (0.132) [0.114, 0.185] | 1.95 (0.146) [0.107, 0.178] | -10.0563 [-19.868, 29.394] |
| Anterior cingulate gyrus | 10.91 (0.816) [0.679, 1.070] | 5.43 (0.407) [0.303, 0.551] | 5.48 (0.410) [0.338, 0.558] | -0.8096 [-37.740, 26.895] |
| Middle cingulate gyrus | 9.58 (0.717) [0.657, 0.929] | 4.95 (0.371) [0.320, 0.480] | 4.62 (0.346) [0.308, 0.478] | 6.8771 [-22.671, 26.306] |
| Posterior cingulate gyrus | 8.19 (0.613) [0.574, 0.816] | 4.04 (0.302) [0.266, 0.411] | 4.15 (0.311) [0.292, 0.420] | -2.8148 [-23.339, 13.354] |
| Parahippocampal gyrus | 5.20 (0.389) [0.342, 0.509] | 2.43 (0.182) [0.160, 0.250] | 2.76 (0.207) [0.174, 0.266] | -12.7341 [-24.324, 10.718] |
| **Insular cortex** | 26.51 (1.984) [1.944, 2.508] | 13.69 (1.024) [0.987, 1.291] | 12.82 (0.959) [0.940, 1.232] | 6.5511 [-4.365, 13.762] |
| Anterior insula | 7.93 (0.594) [0.527, 0.753] | 4.04 (0.303) [0.258, 0.377] | 3.89 (0.291) [0.264, 0.381] | 3.8298 [-12.708, 9.423] |
| Posterior insula | 4.17 (0.312) [0.270, 0.405] | 2.20 (0.165) [0.135, 0.210] | 1.97 (0.147) [0.130, 0.200] | 11.0971 [-12.441, 21.147] |
| Central operculum | 7.87 (0.589) [0.511, 0.701] | 3.88 (0.290) [0.245, 0.357] | 3.99 (0.299) [0.251, 0.359] | -2.8346 [-20.837, 18.047] |
| Frontal operculum | 4.26 (0.319) [0.238, 0.383] | 1.90 (0.142) [0.111, 0.190] | 2.36 (0.176) [0.114, 0.206] | -21.2908 [-38.308, 27.798] |
| Parietal operculum | 2.27 (0.170) [0.240, 0.422] | 0.79 (0.059) [0.094, 0.196] | 1.48 (0.111) [0.131, 0.241] | -60.3488 [-61.341, 11.337] |

| Region | | | | |
|---|---|---|---|---|
| Sup. frontal gyrus | 2.43 (0.022) [0.016, 0.024] | 2.50 (0.023) [0.016, 0.024] | 2.37 (0.021) [0.016, 0.024] | 5.3994 [-9.489, 10.788] |
| Sup. frontal gyrus medial segment | 2.97 (0.027) [0.022, 0.031] | 3.06 (0.028) [0.022, 0.032] | 2.89 (0.026) [0.021, 0.030] | 5.8173 [-10.125, 20.782] |
| Supplementary motor cortex | 2.56 (0.023) [0.018, 0.027] | 2.64 (0.024) [0.018, 0.026] | 2.49 (0.023) [0.017, 0.027] | 5.9940 [-12.970, 18.991] |
| **Parietal lobe** | 2.92 (0.027) [0.024, 0.032] | 2.97 (0.027) [0.024, 0.033] | 2.88 (0.026) [0.023, 0.032] | 2.9251 [-4.783, 8.089] |
| Angular gyrus | 2.09 (0.019) [0.017, 0.026] | 2.09 (0.019) [0.017, 0.026] | 2.09 (0.019) [0.016, 0.026] | 0.0094 [-19.560, 17.275] |
| Postcentral gyrus | 1.34 (0.012) [0.009, 0.015] | 1.35 (0.012) [0.009, 0.014] | 1.34 (0.012) [0.009, 0.015] | 0.2877 [-23.779, 15.648] |
| Postcentral gyrus medial segment | 1.03 (0.009) [0.005, 0.014] | 0.84 (0.008) [0.004, 0.015] | 1.17 (0.011) [0.005, 0.014] | -32.6647 [-47.525, 48.303] |
| Precuneus | 2.18 (0.020) [0.017, 0.028] | 2.15 (0.020) [0.017, 0.028] | 2.22 (0.020) [0.018, 0.028] | -3.0447 [-16.186, 9.807] |
| Sup. parietal lobule | 1.37 (0.012) [0.009, 0.017] | 1.34 (0.012) [0.009, 0.016] | 1.40 (0.013) [0.010, 0.018] | -4.3384 [-25.711, 11.491] |
| Supramarginal gyrus | 2.30 (0.021) [0.017, 0.024] | 2.22 (0.020) [0.016, 0.024] | 2.38 (0.022) [0.018, 0.025] | -7.2624 [-19.914, 15.633] |
| **Temporal lobe** | 1.80 (0.016) [0.014, 0.021] | 1.82 (0.017) [0.014, 0.022] | 1.77 (0.016) [0.014, 0.021] | 2.5423 [-6.419, 10.867] |
| Fusiform gyrus | 3.23 (0.029) [0.026, 0.037] | 3.20 (0.029) [0.025, 0.037] | 3.27 (0.030) [0.026, 0.037] | -1.9729 [-11.909, 8.787] |
| Planum polare | 1.66 (0.015) [0.012, 0.022] | 1.77 (0.016) [0.012, 0.022] | 1.58 (0.014) [0.012, 0.023] | 11.0179 [-28.235, 24.077] |
| Planum temporale | 1.76 (0.016) [0.015, 0.024] | 1.76 (0.016) [0.014, 0.024] | 1.77 (0.016) [0.015, 0.025] | -0.4174 [-30.530, 22.929] |
| Inf. temporal gyrus | 3.20 (0.029) [0.024, 0.035] | 3.10 (0.028) [0.023, 0.035] | 3.29 (0.030) [0.025, 0.036] | -6.0325 [-15.086, 5.983] |
| Middle temporal gyrus | 2.87 (0.026) [0.023, 0.032] | 2.80 (0.025) [0.023, 0.032] | 2.94 (0.027) [0.023, 0.033] | -4.8303 [-12.170, 9.294] |
| Sup. temporal gyrus | 2.25 (0.020) [0.019, 0.027] | 2.25 (0.020) [0.019, 0.027] | 2.25 (0.020) [0.019, 0.028] | 0.2740 [-18.296, 13.652] |
| Transverse temporal gyrus | 1.72 (0.016) [0.014, 0.024] | 1.83 (0.017) [0.013, 0.024] | 1.64 (0.015) [0.013, 0.025] | 11.1110 [-26.784, 23.122] |
| Temporal pole | 3.50 (0.032) [0.027, 0.038] | 3.45 (0.031) [0.026, 0.038] | 3.54 (0.032) [0.026, 0.039] | -2.6149 [-12.478, 13.248] |
| **Occipital lobe** | 1.83 (0.017) [0.014, 0.024] | 1.85 (0.017) [0.014, 0.024] | 1.81 (0.016) [0.014, 0.024] | 2.1961 [-7.260, 10.243] |
| Calcarine cortex | 1.05 (0.010) [0.007, 0.021] | 1.07 (0.010) [0.007, 0.021] | 1.03 (0.009) [0.007, 0.021] | 4.0774 [-26.722, 23.937] |
| Cuneus | 1.15 (0.010) [0.009, 0.020] | 1.14 (0.010) [0.009, 0.020] | 1.16 (0.011) [0.009, 0.021] | -1.6852 [-23.282, 21.935] |
| Lingual gyrus | 2.20 (0.020) [0.016, 0.028] | 2.11 (0.019) [0.015, 0.028] | 2.28 (0.021) [0.016, 0.028] | -7.6034 [-15.088, 11.006] |
| Occipital fusiform gyrus | 2.13 (0.019) [0.015, 0.026] | 2.11 (0.019) [0.015, 0.027] | 2.16 (0.020) [0.014, 0.026] | -2.0759 [-16.149, 24.382] |
| Inf. occipital gyrus | 2.20 (0.020) [0.016, 0.026] | 2.12 (0.019) [0.016, 0.025] | 2.28 (0.021) [0.016, 0.027] | -7.0097 [-19.724, 11.927] |
| Middle occipital gyrus | 2.27 (0.021) [0.017, 0.028] | 2.19 (0.020) [0.017, 0.028] | 2.35 (0.021) [0.017, 0.029] | -7.0092 [-21.709, 16.514] |
| Sup. occipital gyrus | 1.37 (0.012) [0.009, 0.019] | 1.45 (0.013) [0.009, 0.019] | 1.29 (0.012) [0.009, 0.019] | 11.6807 [-22.524, 29.472] |
| Occipital pole | 0.95 (0.009) [0.006, 0.018] | 1.06 (0.010) [0.006, 0.018] | 0.84 (0.008) [0.006, 0.018] | 22.9264 [-32.332, 32.801] |
| **Limbic cortex** | 2.88 (0.026) [0.024, 0.037] | 2.88 (0.026) [0.023, 0.037] | 2.88 (0.026) [0.024, 0.033] | 0.0460 [-8.343, 5.104] |
| Entorhinal area | 3.23 (0.029) [0.025, 0.033] | 3.20 (0.029) [0.025, 0.034] | 3.26 (0.030) [0.024, 0.033] | -1.7133 [-12.898, 21.043] |
| Anterior cingulate gyrus | 3.25 (0.030) [0.026, 0.037] | 3.21 (0.029) [0.026, 0.038] | 3.29 (0.030) [0.026, 0.037] | -2.5043 [-9.450, 13.544] |
| Middle cingulate gyrus | 2.83 (0.026) [0.021, 0.032] | 2.85 (0.026) [0.022, 0.033] | 2.80 (0.025) [0.020, 0.032] | 1.9931 [-10.849, 16.955] |
| Posterior cingulate gyrus | 2.62 (0.024) [0.022, 0.033] | 2.60 (0.024) [0.021, 0.033] | 2.63 (0.024) [0.022, 0.033] | -0.9348 [-12.663, 8.929] |
| Parahippocampal gyrus | 2.38 (0.022) [0.019, 0.029] | 2.43 (0.022) [0.019, 0.029] | 2.33 (0.021) [0.019, 0.029] | 4.1444 [-13.146, 17.107] |
| **Insular cortex** | 2.61 (0.024) [0.021, 0.030] | 2.58 (0.023) [0.021, 0.030] | 2.65 (0.024) [0.021, 0.030] | -2.4455 [-10.197, 6.570] |
| Anterior insula | 3.23 (0.029) [0.026, 0.035] | 3.30 (0.030) [0.026, 0.036] | 3.17 (0.029) [0.025, 0.035] | 4.1467 [-7.383, 12.108] |
| Posterior insula | 2.47 (0.022) [0.019, 0.031] | 2.52 (0.023) [0.019, 0.031] | 2.40 (0.022) [0.018, 0.031] | 4.9855 [-13.725, 19.672] |
| Central operculum | 2.31 (0.021) [0.019, 0.029] | 2.23 (0.020) [0.019, 0.029] | 2.39 (0.022) [0.019, 0.029] | -6.8677 [-15.241, 15.275] |
| Frontal operculum | 2.60 (0.024) [0.021, 0.031] | 2.62 (0.024) [0.020, 0.032] | 2.59 (0.024) [0.020, 0.031] | 0.8353 [-13.813, 19.227] |
| Parietal operculum | 1.78 (0.016) [0.015, 0.025] | 1.75 (0.016) [0.015, 0.025] | 1.79 (0.016) [0.015, 0.026] | -1.8790 [-27.309, 19.946] |